\documentclass[aps,prb,reprint,superscriptaddress,amsmath,amssymb,floatfix]{revtex4-2}

\usepackage{graphicx}
\usepackage{dcolumn}
\usepackage{bm}
\usepackage{physics}
\usepackage{tikz}
\usepackage{tikz-feynman}
\tikzfeynmanset{compat=1.1.0}
\usepackage{hyperref}
\usepackage{xcolor}
\usepackage{subfig}
\usepackage[singlelinecheck=false,justification=raggedright]{caption}

\hypersetup{
    colorlinks=true,
    citecolor=blue,
    linkcolor=blue,
    urlcolor=blue
}

\begin{document}


\title{Topology and Spectral Entanglement in Cavity-Mediated Photon Scattering}

\author{Eric~R.~Bittner}
\email{ebittner@central.uh.edu}
\affiliation{Department of Physics, University of Houston, Houston, Texas 77204, United~States}

\author{Andrei Piryatinski}
\email{apiryat@lanl.gov}
\affiliation{Theoretical Division, Los Alamos National Laboratory, Los Alamos, NM 87545, United~States }


\date{\today}

\begin{abstract}

We develop a diagrammatic theory of cavity-mediated photon-photon interactions in a topological insulator using the SSH model. The fourth-order vertex $\Gamma^{(4)}(\omega_1,\omega_2)$ governs spectral entanglement and Kerr nonlinearity, leading to a nonlinear topological phase diagram. We also compute the electronic self-energy from vacuum photon exchange and identify symmetry-imposed limits on band renormalization. These results link band geometry to emergent light-matter correlations.
\end{abstract}

\maketitle


\section{Introduction}

The emergence of topological band structures in condensed matter systems has prompted parallel developments in photonics, where topological effects can be engineered in photonic lattices, metamaterials, and strongly coupled light--matter platforms\cite{Ozawa2019}. A major goal in this direction is to exploit band topology to control optical response---particularly in the nonlinear or quantum regime, where few-photon states interact via a structured medium\cite{Carusotto2013, Ciuti2005}.

Topological photonic systems based on chiral edge modes\cite{Klembt2018, Gianfrate2020}, polaritonic lattices\cite{Li2019}, or synthetic gauge fields have exhibited robustness to disorder and directionality in transport, but the role of topology in shaping \emph{nonlinear} photon--photon interactions remains less well understood. In standard QED, such interactions arise from virtual electron--positron pairs and are described by polarization loops and box diagrams\cite{Heisenberg1936, Euler1936}. Analogous effects can occur in solid-state cavity QED, where photon--photon scattering is mediated by virtual excitations of the medium\cite{Imamoglu1997, Hartmann2006}. Here, the structure of the medium---including its band topology---can imprint on the nonlinear interaction vertex.

In this work, we present a microscopic quantum electrodynamical theory for cavity-mediated photon--photon interactions in a one-dimensional topological insulator. Using the Su--Schrieffer--Heeger (SSH) model as a tunable photonic medium, we compute the fourth-order interaction vertex \( \Gamma^{(4)}(\omega_1, \omega_2;\omega_3,\omega_4) \) that governs biphoton scattering
$(\omega_1,\omega_2)_{in}\to (\omega_3,\omega_4)_{out}$. This vertex is constructed diagrammatically from polarization bubbles and electron--hole interactions, and includes both resonant and off-resonant contributions.

We show that the effective photon--photon interaction strength and Kerr nonlinearity are strongly influenced by the topological phase, while the spectral entanglement of the biphoton state is governed primarily by local band geometry and momentum-space coherence. By applying a stationary phase approximation and Schmidt decomposition, we identify the dominant contributions to the scattering vertex and analyze how curvature and dipole structure modulate the degree of non-separability. Our results clarify the connection between band structure and photon--photon correlations, and distinguish between topological signatures in linear versus nonlinear optical response.

\section{Theoretical
Model}

We consider a model in which a single quantized mode of an optical cavity is coupled to a 1D electronic system described by the Su-Schrieffer–Heeger (SSH) Hamiltonian\cite{SSH1979}.
The model consists of a bipartite chain with alternating hopping amplitudes \( t_1 \) and \( t_2 \), connecting adjacent lattice sites of sublattices \( A \) and \( B \).
Here it provides a useful tight-binding model exhibiting 
a topological phase. 
Briefly, when $t_1>t_2$, the system is in a trivial phase, when $t_1=t_2$ the system is
in a metallic state, 
and when $t_1<t_2$, the system enters a topological phase at half-filling. 
The details of this model are provided in the Appendix. 
The light-matter coupling is taken to be in the velocity gauge, where the light–matter interaction enters via a current operator $\hat{\jmath}$. The total Hamiltonian is written as
\begin{equation}
H = H_{\text{SSH}} + \omega_c \hat{a}^\dagger \hat{a} + g \hat{\jmath} \cdot (\hat{a} + \hat{a}^\dagger),
\end{equation}
where $\omega_c$ is the bare cavity frequency and $g$ is the light–matter coupling constant. {The choice of velocity gauge ensures that the light–matter interaction is mediated by the electronic current rather than the dipole operator, which is essential for systems with nontrivial band topology.} 
The current operator $\hat{\jmath}$ is defined as the $k$-space velocity matrix element between valence and conduction bands:
\begin{equation}
\hat{\jmath} \sim \sum_k \mu(k)\, c_{c,k}^\dagger c_{v,k} + \text{H.c.},
\end{equation}
where $\mu(k) = \langle u_{c,k} | \partial_k H_{\text{SSH}} | u_{v,k} \rangle$ represents the interband current matrix element. \textcolor{black}{This form correctly accounts for interband transitions induced by cavity photons.}
Using the explicit form of the Bloch wavefunctions and the connection between the position operator and momentum derivatives in periodic systems \cite{Blount1962, Resta1994}, we find:
\begin{equation}
\mu(k) = \frac{ t_1 t_2 \sin k}{\Delta(k)}.
\label{eq:dipole}
\end{equation}
This vanishes at the Brillouin zone boundaries (\( k = 0, \pi \)) and changes sign across the zone, reflecting the inversion symmetry of the chain. It reaches its maximum magnitude when \( k \approx \pi/2 \) and \( t_1 \sim t_2 \).
We note that the magnitude and sign of \( \mu(k) \) play a crucial role in determining both the linear optical absorption and the nonlinear photon–photon interaction vertex.

\subsection{Cavity self-energy}
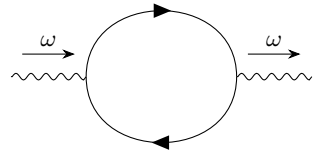
\begin{figure}[th]
\centering
\begin{tikzpicture}
\begin{feynman}
  \vertex (i) at (-2,0);
  \vertex (f) at (2,0);
  \vertex (l) at (-1,0);
  \vertex (r) at (1,0);
  
  \vertex (t) at (0,0.8);
  \vertex (b) at (0,-0.8);

  \diagram*{
    (i) -- [photon, momentum=\(\omega\)] (l),
    (r) -- [photon, momentum=\(\omega\)] (f),
    (l) -- [fermion, half left] (r),
    (r) -- [fermion, half left] (l),
  };
\end{feynman}
\end{tikzpicture}
\caption{Second-order photon self-energy diagram. A cavity photon creates a virtual electron–hole pair in the SSH chain, which then recombines to emit a photon. This renormalizes the cavity photon mode via a polarization bubble, encoding the spectral features of the underlying topological medium.}
\label{fig:photon_self_energy}
\end{figure}

When an optical cavity mode couples to interband transitions in a material, the cavity photon acquires a self-energy due to virtual electron–hole pair excitations within the 
medium with in the cavity. 
Using the Feschbach method
we can formally solve the 
time-independent Schr\"odinger
equation for the material 
as driven by the cavity
and reintroduce this into the 
Schr\"odinger equation for 
the cavity (photon) wavefunction.
\begin{align}
    \left(H_{PP} - E - \Sigma^{R}
    (\omega)\right)\psi_P &= 0. \\
\label{eq:sigma2}
\end{align}
Here we have defined the (retarded) 
photon self-energy
\begin{equation}
\Sigma^{R}(\omega) = \int_{\text{BZ}} \frac{dk}{2\pi} \frac{|\mu(k)|^2}{\omega - \Delta(k) + i\eta},
\label{eq:sigma2b}
\end{equation}
\textcolor{black}{Here, the integration is over the electronic crystal momentum $k$ within the Brillouin zone; this is \emph{distinct} from the cavity photon wavevector, which is typically centered around $q=0$ and governed by the cavity geometry.} \textcolor{black}{The coupling arises because the photon couples collectively to electronic transitions across the full band structure.}
In the present context, \( \Sigma^{R}(\omega) \) arises from virtual electron–hole excitations within the SSH chain and reflects both the topological phase and the spectral structure of the electronic transitions.
The Feynman diagram for this 
is shown in Fig.~\ref{fig:photon_self_energy}
where an input photon is 
annihilated to create a
virtual electron/hole pair 
at momentum $k$, which in 
turn undergoes 	annihilation 
to create the outgoing 
photon. Since this is 
formally an energy conserving process, the input and output 
photons have the same energy.
The integration over the 
complete Brillouin zone 
arises from inserting a complete
set of virtual electron/hole states.
This self-energy reflects the renormalization of the cavity photon mode due to its coupling to the interband continuum. The real part leads to a shift in resonance frequency, while the imaginary part encodes photon absorption (decay) into the electron–hole continuum. 

The dressed photon propagator becomes:
\begin{equation}
G_{\text{cav}}^{R}(\omega) = \frac{1}{\omega - \omega_c(q) - \Sigma^{R}(\omega)},
\end{equation}
where \( \omega_c(q) = \omega_c(0) + \beta q^2 \) is the bare cavity dispersion and \( \beta \) is the effective photon mass parameter. For \( \omega \approx \omega_c \sim \Delta_0 \), resonant hybridization leads to polariton formation.
For consistency, we will use $k$ to denote the crystal momentum
and $q$ to denote the cavity photon
momentum in the direction in 
the $xy$ plane of the cavity. 

This expression is central to modeling the cavity transmission spectrum or local density of states (LDOS), as the imaginary part of \( G_{\text{cav}}^R(\omega, q) \) gives direct access to the absorptive and emissive properties of the system:
\begin{equation}
\mathcal{A}(\omega, q) = -\frac{1}{\pi} \Im G_{\text{cav}}^R(\omega, q),
\label{eq:13}
\end{equation}
which is proportional to the experimentally measurable transmission or reflection coefficient of the cavity under weak excitation.

\begin{figure*}[th]
    \centering
\subfloat[]{\includegraphics[width=0.4\textwidth]{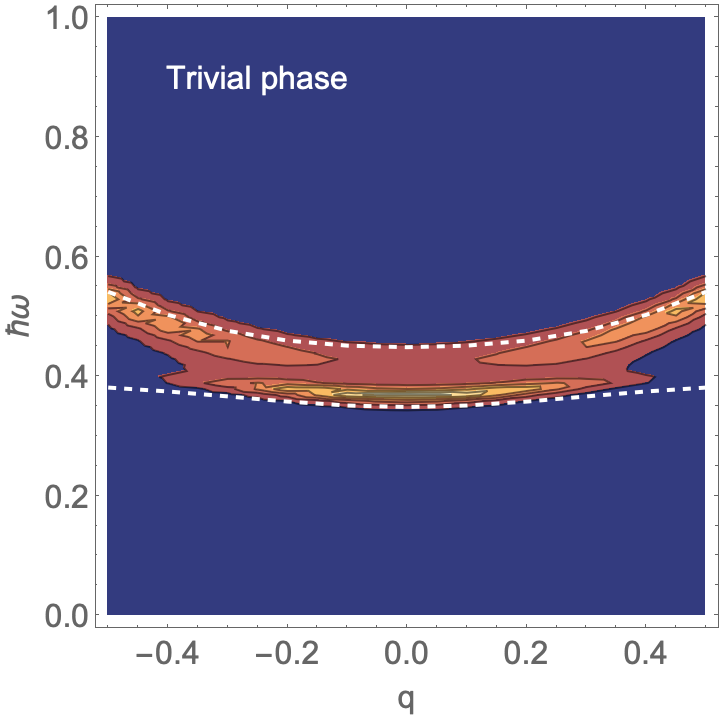}}
\subfloat[]{\includegraphics[width=0.4\textwidth]{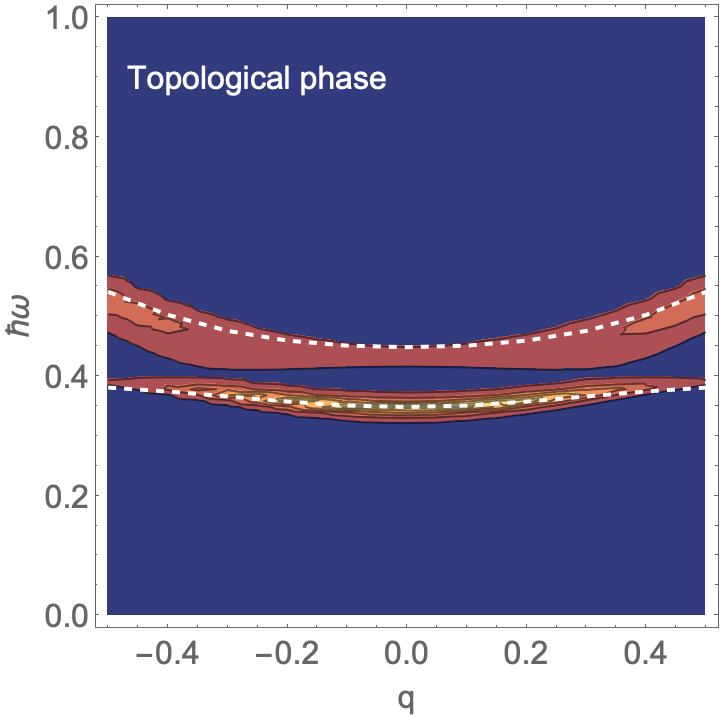}}

    \caption{Polariton dispersion relations computed from the dressed cavity Green’s function $G_\mathrm{ph}^R(\omega, q)$ for (a) the trivial phase $(t_1 > t_2)$ and (b) the topological phase $(t_2 > t_1)$. The cavity frequency is set at resonance with the optical transition: $\omega_c = 2|t_1 - t_2|$. Superimposed are the upper and lower polariton branches obtained via exact diagonalization of a simple Hopfield Hamiltonian for a cavity resonant with the band-gap at $k=\pi$. 
        }
    \label{fig:dispersion}
\end{figure*}

Fig.~\ref{fig:dispersion} shows the absorption spectrum (Eq.~\ref{eq:13}) for two cases with identical band gaps; however, one (A) is in the trivial phase
in which $t_2< t_1$, while in (B)
we set $t_2>t_1$ and set the 
cavity frequency $\omega_c$ to be 
equal to the band-gap of the
SSH lattice $\omega_c = 2|t_1-t_2|$.   First, we note that 
Eq.~\ref{eq:13} faithfully reproduces the lower and
upper polariton branches of the system as indicated by the dashed curves superimposed on the figure. 
To see this, we have superimposed the dispersion curves for the much simpler 
Hopfield model 
\begin{align}
H_{Hopfield} = \left[
\begin{array}{cc}
 \beta q^{2} + \Delta(\pi)  & g \\
g  &  \Delta(\pi)
\end{array}
\right] 
\end{align}
where $g = 0.05$, $\beta = 0.5$, and $q$ is the cavity wavevector.  Here, we have adjusted $g$ to reproduce the upper and lower polariton 
splitting in the topological phase.  
However, the self-energy term contains information about the electronic structure of the entire 
material subsystem.  This is reflected in the relative shift 
of the absorption spectrum at $q=0$. It is tempting to 
attribute this shift to 
the underlying change in the
geometric topology of the 
SSH Bloch states. If this were the 
case, one would expect to 
see a rotation in the overall
phasing of the full linear response. Such shifts in phase
can occur in the non-linear
response, as we have recently
shown where by the real and 
imaginary components of the 
response acquire a  $\pi/2$ 
phase shift close to $\omega=0$ (non-linear DC response) when the 
system is in the topological 
regime.

To determine the nonlinear optical properties of the cavity, we solve the self-consistent equation
\begin{equation}
\omega_n = \omega_c + \Sigma^{R}(\omega_n, n),
\label{eq:omega_n}
\end{equation}
for a discrete set of photon numbers $n = 0, 1, 2, \ldots$, yielding the $n$-dependent cavity resonance frequencies $\omega_n$ 
and self-energy
\begin{equation}
\Sigma^{R}(\omega, n) = g^2 (n + 1) \int_{-\pi}^{\pi} \frac{dk}{2\pi} \, \frac{|\mu(k)|^2}{\omega - \Delta(k) + i\eta},
\label{eq:Sigma_n}
\end{equation}
where $g$ is the light–matter coupling constant and the factor $(n+1)$ accounts for stimulated and spontaneous contributions.
 The mean-field photon–photon interaction strength can then be extracted from a Taylor expansion of $\omega_n$:
\begin{equation}
\omega_n \approx \omega_0 + U n + \frac{1}{2} U' n^2 + \cdots,
\label{eq:omega_expansion}
\end{equation}
where $U$ is the Kerr interaction coefficient, and $U'$ encodes higher-order nonlinearities. In the weak coupling regime, a closed-form expression for $U$ can be obtained by differentiating Eq.~\eqref{eq:Sigma_n} with respect to $n$:
\begin{equation}
U = g^2 \int_{-\pi}^{\pi} \frac{dk}{2\pi} \, \frac{|\mu(k)|^2}{[\omega_c - \Delta(k) + i\eta]^2}.
\label{eq:Kerr_U}
\end{equation}

\begin{figure}[th]
\centering
\includegraphics[width=0.45\textwidth]{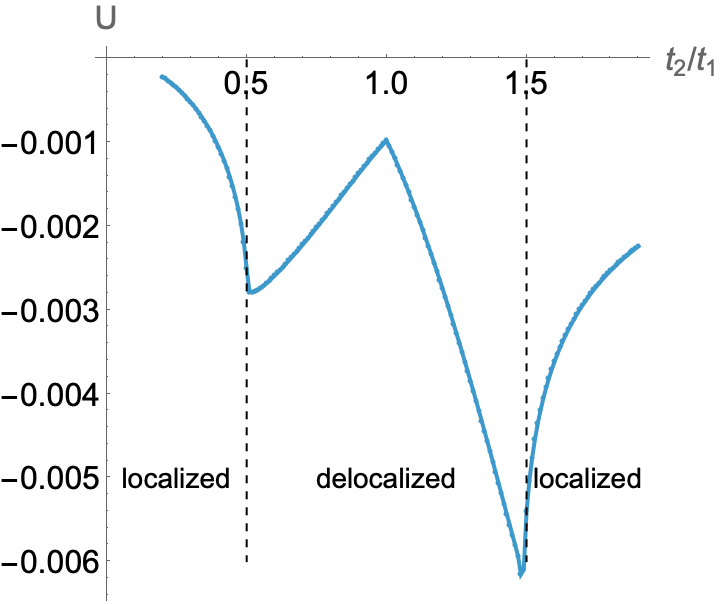}
\caption{
Kerr interaction coefficient \(U\) extracted from self-consistent solutions to the photon self-energy equation, plotted as functions of the SSH dimerization ratio \(t_2/t_1\).
Vertical dashed lines denote localized vs. delocalized regimes.
}
\label{fig:4}
\end{figure}

We numerically solve the self-consistent equation \eqref{eq:omega_n} to extract $\omega_n$ and fit the resulting nonlinear dispersion to determine $U$  as a function of the SSH parameter $r = t_2/t_1$. This provides a means to explore how topological band structure affects effective photon–photon interactions in a microcavity geometry.
Results are shown in Fig~\ref{fig:4}.

The sharp variation of the Kerr interaction strength \( U \) near the critical point \( t_1 = t_2 \) is reminiscent of the Peierls instability in one-dimensional metals~\cite{Peierls1955}. In the Peierls picture, a uniform chain with a half-filled band is unstable to a lattice distortion that doubles the unit cell, opens a gap at the Brillouin zone boundary, and lowers the system’s total energy. The SSH model similarly encodes dimerization through alternating hopping amplitudes \( t_1 \) and \( t_2 \), and undergoes a topological phase transition when the stronger bond shifts from intra-cell to inter-cell. Although this transition is formally topological, characterized by a change in the Zak phase, it retains many features of symmetry breaking, including a two-fold degenerate ground state and solitonic domain walls.

This analogy is especially vivid in the context of conjugated polymers such as polyacetylene, where mid-gap states emerge at domain boundaries separating distinct dimerization patterns~\cite{Heeger1988}. In our cavity-coupled system, the nonlinear optical response of the photon field is governed by the photon self-energy, which is mediated by interband electronic transitions within the SSH lattice. As \( t_2 \to t_1 \), the band gap closes and the polarizability diverges, leading to an enhancement—and eventual sign change—of the Kerr interaction. This mirrors the enhanced susceptibility and soft-mode behavior that accompany the onset of a Peierls transition~\cite{Gruner1994}. Our results therefore suggest that a strongly coupled cavity can serve as a sensitive spectroscopic probe of Peierls-type instabilities, even in systems whose electronic structure supports topologically nontrivial phases.
Recent work has further extended Peierls physics into cold-atom~\cite{Atala2013}, ultrafast~\cite{Rettig2016}, Floquet-driven~\cite{Hubener2017}, and photonic~\cite{Mukherjee2020} platforms, suggesting that cavity-based nonlinear optics may offer a complementary route to probing symmetry breaking and collective dynamics in quantum materials.
These features highlight the sensitivity of cavity electrodynamics to the underlying topological phase of the embedded material, and provide a direct, linear-optical probe of topological order. Unlike traditional transport measurements, the photonic Green's function can be accessed via spectroscopy and does not require contacts or local gating, making it especially valuable in probing optically active topological materials.

\subsection{Keldyshh Formalism and Nonequilibrium Photon Observables}

In order to capture the nonequilibrium dynamics and spectral properties of the cavity field in our SSH–cavity system, it is instructive to reframe the theory using the Keldysh path integral formalism\cite{kamenev2011}. Unlike the equilibrium Matsubara framework, the Keldysh approach allows us to analyze time-dependent phenomena, dissipation, and quantum noise within a unified language. This is particularly relevant for optical cavities, which are inherently open quantum systems subject to loss, drive, and decoherence. In what follows, we construct the effective action for the photon field by integrating out the electronic degrees of freedom in the SSH chain, and we demonstrate how the resulting Green’s functions encode measurable quantities such as the cavity absorption spectrum and frequency-resolved photon occupation.

\begin{figure}[h]
  \centering
  \includegraphics[width=\columnwidth]{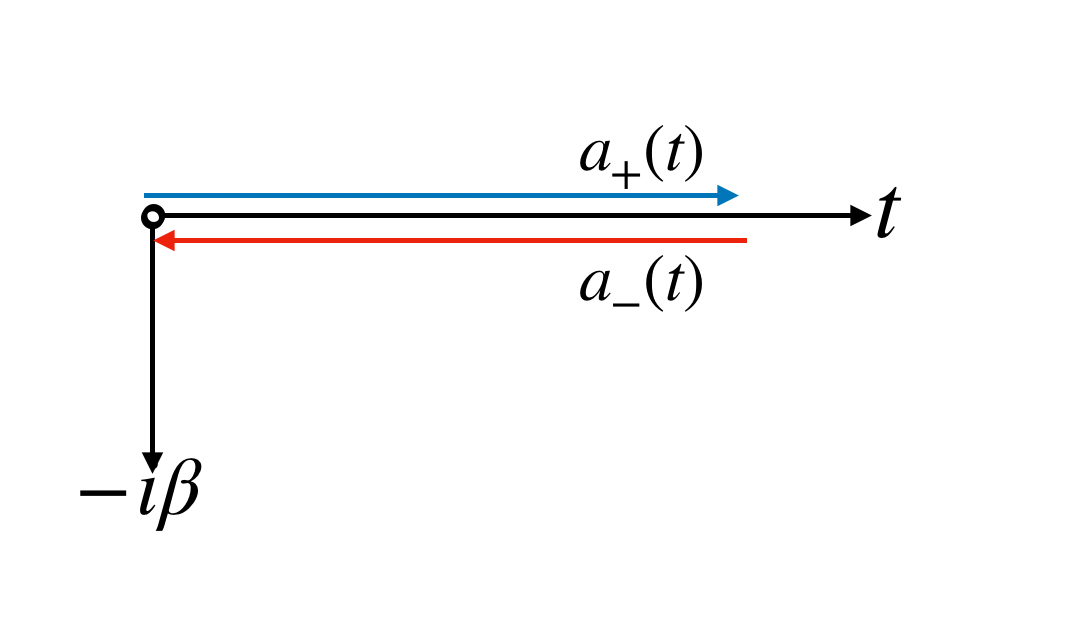}
  \caption{Illustration of the Keldysh time contour $\mathcal{C}$ used to formulate nonequilibrium dynamics. The system evolves forward in real time from $t_o$ to $t_{\mathrm{f}}$, then backward from $t_{\mathrm{f}}$ to $t_o$, and (optionally) along an imaginary-time branch to prepare an initial thermal state at inverse temperature $\beta$.}
  \label{fig:keldysh-contour}
\end{figure}

We begin by defining the cavity photon field on the closed time contour, which consists of a forward $(+)$ branch and a backward $(-)$ branch as illustrated in 
Fig.~\ref{fig:keldysh-contour}.  The fields on these branches, $a_+(t)$ and $a_-(t)$, are used to generate correlation functions via path integration. To streamline the treatment of real-time dynamics, we perform a Keldysh rotation to define the classical and quantum components of the field:
\begin{equation}
a_{\mathrm{cl}}(t) = \frac{a_+(t) + a_-(t)}{\sqrt{2}}, \quad
\, a_{\mathrm{q}}(t) = \frac{a_+(t) - a_-(t)}{\sqrt{2}},
\end{equation}
where $a_{\mathrm{cl}}$ represents the average field and $a_{\mathrm{q}}$ encodes quantum fluctuations. The effective action in this basis takes the general quadratic form:
\begin{widetext}
\begin{equation}
S_{\mathrm{eff}}[a_{\mathrm{cl}}, a_{\mathrm{q}}] = 
\int \frac{d\omega}{2\pi}
\begin{pmatrix}
a_{\mathrm{cl}}^*(\omega) & a_{\mathrm{q}}^*(\omega)
\end{pmatrix}
\begin{pmatrix}
0 & [G^A(\omega)]^{-1} \\
[G^R(\omega)]^{-1} & G^K(\omega)
\end{pmatrix}
\begin{pmatrix}
a_{\mathrm{cl}}(\omega) \\
a_{\mathrm{q}}(\omega)
\end{pmatrix},
\end{equation}
where $G^R$, $G^A$, and $G^K$ are the retarded, advanced, and Keldysh Green’s functions of the cavity field, respectively. These functions are determined by the underlying electron–hole polarization processes in the SSH chain. Details of 
the derivation of our equations
are given in the Appendix.
\end{widetext}
The retarded photon propagator is obtained from Dyson resummation:
\begin{equation}
G^R(\omega) = \frac{1}{\omega - \omega_c - \Sigma^R(\omega) + i\eta},
\end{equation}
where $\omega_c$ is the bare cavity frequency, and $\Sigma^R(\omega)$ is the retarded photon self-energy due to coupling with the SSH chain. This self-energy is computed microscopically from the polarization bubble:
\begin{equation}
\Sigma^R(\omega) = g^2 \int_{-\pi}^{\pi} \frac{|\mu(k)|^2}{\omega - \Delta(k) + i\eta} \, \frac{dk}{2\pi},
\end{equation}
with $\Delta(k)$ the interband energy gap and $\mu(k)$ the dipole matrix element coupling the cavity field to the SSH electrons. The shape and magnitude of $\Sigma^R(\omega)$ are directly influenced by the curvature of the SSH band structure and the distribution of $\mu(k)$, both of which vary with the topological phase.

From the retarded propagator, we define the photon spectral function:
\begin{align}
\mathcal{A}(\omega) &= -\frac{1}{\pi} \, \mathrm{Im} \, G^R(\omega), \\
\label{eq:KeldA}
\end{align}
This spectral function characterizes the energy-dependent response of the cavity to an external probe and exhibits key signatures of strong coupling. 
Importantly, it is identical to 
the spectral function we obtained
in Eq.~\ref{eq:13} using the 
Feshbach approach.

To obtain the photon occupation spectrum, we examine the Keldysh component $G^K(\omega)$, which encodes the symmetrized correlation function $\langle \{a(\omega), a^\dagger(\omega)\} \rangle$. In the presence of thermal noise or quantum fluctuations, the Keldysh self-energy satisfies the relation
\begin{equation}
\Sigma^K(\omega) \approx -2i\, \mathrm{Im}\, \Sigma^R(\omega)\, [1 + 2n_B(\omega)],
\end{equation}
where $n_B(\omega)$ is the Bose--Einstein distribution evaluated at the electronic temperature. The full Keldysh Green’s function is then given by
\begin{equation}
G^K(\omega) = G^R(\omega)\, \Sigma^K(\omega)\, G^A(\omega),
\end{equation}
and from this we define the energy-resolved photon occupation:
\begin{equation}
n(\omega) = \frac{1}{2} \left( \frac{G^K(\omega)}{2i\, \mathrm{Im}\, G^R(\omega)} - 1 \right).
\end{equation}
This expression generalizes the fluctuation--dissipation theorem to nonequilibrium settings and enables a microscopic calculation of steady-state photon statistics.

The Keldysh formalism thus provides a comprehensive theoretical framework to describe both the spectral and statistical properties of the cavity field in the presence of a structured, correlated electronic environment. Importantly, it allows for systematic inclusion of dissipation, quantum noise, and time-dependent drive, all of which play essential roles in modern cavity QED and photonic materials platforms. By embedding the SSH chain into this formalism, we create a foundation for studying driven--dissipative topological systems, nonlinear spectroscopy, and quantum light–matter interfaces beyond the perturbative regime.

This derivation ensures consistency between the Feshbach method and the  Keldysh nonequilibrium theory, since
both routes give the same self-energy which govern the linear response and spectral functions.
Hence, the Feshbach approach  faithfully reproduces the photon Green’s function, including all renormalizations due to virtual electron--hole excitations in the medium. 

\subsection{Vacuum Fluctuation--Induced Band Mixing: One-Loop Electronic Self-Energy}
To characterize how vacuum fluctuations of the cavity field reshape the electronic structure of the SSH chain, we compute the one-loop fermionic self-energy induced by virtual photon exchange. Even at zero temperature, these quantum fluctuations mediate virtual transitions that hybridize valence and conduction bands, leading to momentum-resolved dressing of the electronic quasiparticles. This dressing modifies both the dispersion and coherence properties of the fermions, and introduces a new channel for cavity-mediated electronic structure control. The one-loop self-energy provides a natural starting point for quantifying these effects, allowing us to derive an effective band Hamiltonian that incorporates the influence of the cavity field in a systematic and analytically transparent way.

We start from the light--matter interaction Hamiltonian in momentum space which we recall from the previous section,
\begin{equation}
H_{\mathrm{int}} = g \sum_k \mu(k) \left[ a \, c_k^\dagger v_k + a^\dagger \, v_k^\dagger c_k \right],
\end{equation}
which couples valence and conduction band states via absorption or emission of a cavity photon with mode operator $a$. Here, $c_k$ and $v_k$ denote the annihilation operators for conduction and valence band states, respectively, in the diagonal basis of the SSH model.
The quantities $\epsilon_c(k)$ and $\epsilon_v(k)$ used in this section refer to the band energies of the SSH model in its \emph{diagonalized} representation as discussed above 
and in the Appendix.

Whereas above, we considered the self-energy of the cavity in ther presence of the material, here we treat the electrons as the ``system'' and the cavity photons as the ``bath'', and compute the fermionic self-energy to second order in $g$. The leading diagram contributing to the conduction-band self-energy corresponds to emission and reabsorption of a virtual photon:
\begin{equation}
\Sigma^{(c)}(k, \omega) = g^2 \, \mu(k)^2 \, G_{\rm cav}(\omega - \epsilon_v(k)),
\end{equation}
where $\epsilon_v(k)$ is the valence band energy and $G_{\rm cav}^R$ is the retarded Green's function of the cavity photon. Similarly, the (retarded) valence-band self-energy is:
\begin{equation}
\Sigma^{(v)}(k, \omega) = g^2 \, \mu(k)^2 \, G^R_{\rm cav}(\omega - \epsilon_c(k)).
\end{equation}

In both cases, the energy argument reflects the virtual transition between bands mediated by the cavity mode. The bare photon Green's function at zero temperature is given by
\begin{equation}
G^R_{\rm cav}(\omega) = \frac{1}{\omega - \omega_c + i\eta}.
\end{equation}
As a result, the imaginary part of the self-energy is nonzero only when the transition energy matches the cavity mode (on-shell processes). However, the real part induces a Lamb-like shift even off the shell energy given by
\begin{equation}
\mathrm{Re} \, \Sigma^{(c)}(k, \omega) = g^2 \, \mu(k)^2 \, \mathcal{P} \int \frac{d\omega'}{\pi} \, \frac{\mathrm{Im} G^R_{\rm cav}(\omega')}{\omega - \epsilon_v(k) - \omega'},
\end{equation}
where $\mathcal{P}$ denotes the principal value. This expression shows that even in the ground state, the cavity vacuum fluctuations renormalize the electronic dispersion.

In realistic microcavities, the photon mode is not sharply localized at a single frequency, but rather has a finite dispersion due to in-plane momentum structure. This modifies the retarded Green's function to include a momentum-dependent cavity dispersion:
\begin{equation}
G^R_{\rm cav}(q, \omega) = \frac{1}{\omega - \omega_c(q) + i\eta},
\end{equation}
where $\omega_c(q)$ is the photon dispersion, such as $\omega_c(q) = \omega_0 + \alpha q^2$ in a planar cavity geometry. 
The electronic self-energy must then be integrated over all allowed photon momenta:
\begin{equation}
\Sigma^{(c)}(k, \omega) = g^2 \int \frac{dq}{2\pi} \, \mu(k, q)^2 \, G^R_{\rm cav}(q, \omega - \epsilon_v(k)),
\end{equation}
with $\mu(k, q)$ encoding the mode-dependent dipole coupling. This integral accounts for virtual transitions involving all cavity modes, regularizes the ultraviolet behavior of the self-energy, and introduces additional dependence on the electronic band curvature and photon dispersion.

Importantly, since the light--matter interaction is off-diagonal in the electronic band basis, the full fermionic self-energy must be treated as a $2\times2$ matrix. Specifically, in the $(c_k, v_k)$ basis, the fermionic self-energy takes the form:
\begin{equation}
\Sigma^{(f)}(k, \omega) = g^2 \mu(k)^2
\begin{pmatrix}
0 & G^R_{\rm cav}(\omega - \Delta(k)) \\
G^R_{\rm cav}(\omega + \Delta(k)) & 0
\end{pmatrix},
\end{equation}
where $\Delta(k) = \epsilon_c(k) - \epsilon_v(k)$. The cavity thus mediates mixing between valence and conduction bands even at zero temperature.

To obtain the dressed quasiparticle energies, we define the effective Hamiltonian:
\begin{equation}
H^{\text{eff}}(k, \omega) =
\begin{pmatrix}
\epsilon_c(k) & \Sigma^{(f)}_{cv}(k, \omega) \\
\Sigma^{(f)}_{vc}(k, \omega) & \epsilon_v(k)
\end{pmatrix},
\end{equation}
which must be diagonalized at each $k$ and $\omega$ to extract the cavity-dressed band structure. This formalism captures the avoided crossings, level repulsion, and hybridization induced by virtual photon exchange. Even in the weak-coupling regime, the cavity modifies the band edges and introduces nontrivial band curvature, demonstrating how vacuum fluctuations renormalize and mix electronic states.

The eigenvalues of this $2\times2$ matrix can be written analytically as:
\begin{equation}
E_\pm(k, \omega) =  \pm \sqrt{\left(\frac{\epsilon_c(k) - \epsilon_v(k)}{2}\right)^2 + |\Sigma^{(f)}_{cv}(k, \omega)|^2},
\end{equation}
which shows how the dressed energy bands are split by the off-diagonal cavity-mediated coupling. In the limit $|\Sigma^{(f)}_{cv}| \rightarrow 0$, the eigenvalues reduce to the bare bands. In contrast, for finite $\Sigma^{(f)}_{cv}$, the hybridization between valence and conduction bands leads to avoided crossings and modified dispersion, signaling the influence of the vacuum photon field on electronic quasiparticles.

These results also connect to ongoing debates in the literature concerning cavity-mediated modification of chemical reactivity. Experimental studies by Ebbesen and co-workers~\cite{Ebbesen2016AccChemRes, Thomas2016Science} have reported altered reaction rates and ground-state selectivity in molecular ensembles embedded in optical cavities, even under vacuum conditions. These observations have sparked theoretical investigations into whether strong light--matter coupling can reshape Born--Oppenheimer potential energy surfaces or induce new nonadiabatic couplings~\cite{Feist2018NatChem, CenciCohen2023JCP, ZhouYuan2023JPCL}. A central point of contention is whether the cavity coupling strength at the level of an individual molecule is sufficient to significantly modify ground-state properties, or whether observed effects arise from collective phenomena—such as vibrational or electronic polariton formation—or from dynamical processes involving nonequilibrium populations and coherence.

In contrast, the present analysis focuses on a periodic solid-state model in which the photon field couples coherently across the full Brillouin zone. The resulting self-energy corrections are delocalized and momentum-resolved, modifying the band dispersion and inducing interband hybridization through virtual photon exchange. These cavity-mediated effects arise not from localized interactions, but from coherent dressing across the entire electronic structure—a situation more closely analogous to the collective strong coupling regimes explored in polaritonics. This microscopic framework thus provides concrete support for the possibility of cavity-induced electronic modifications, particularly in extended or crystalline systems where collective coupling can be realized. While the implications for molecular chemistry remain under investigation, our results lend theoretical weight to the idea that vacuum fluctuations and light--matter coherence can reshape the electronic landscape when embedded within appropriately engineered photonic environments.

However, a significant
consequence of our analysis is that the off-diagonal electronic self-energy $\Sigma_{cv}(k)$—which mediates interband hybridization through virtual photon exchange—\textbf{vanishes at the Brillouin zone edge}, $k = \pi$. This arises from the symmetry of the SSH model: the dipole matrix element $\mu(k) \sim \sin k$ identically vanishes at $k = \pi$, where the bandgap is minimal and the conduction and valence states are most energetically proximate. As a result, the electronic states near the gap edge—those most relevant for chemical reactivity and low-energy optical transitions—are effectively undressed by the cavity. This implies that, within this model, vacuum fluctuations of the cavity field produce only minimal reorganization of the electronic structure near the Fermi level. The negligible self-energy at $k = \pi$ thus sets a fundamental limit on the capacity of equilibrium cavity QED effects to alter ground-state chemical landscapes in periodic systems of this type. While collective light–matter coupling does renormalize the dispersion away from high-symmetry points, its influence on states directly at the band edge—and thus on possible photochemical or redox processes—may be vanishingly small in the absence of symmetry breaking or cavity mode engineering.

The results presented here are based on a one-loop perturbative analysis in which either the electronic or photonic sector is treated as a passive bath. However, in realistic systems under strong or ultrastrong coupling, a fully self-consistent treatment is required: the photon field dresses the electronic states, and the modified electronic structure in turn feeds back into the photon self-energy. Capturing this mutual renormalization demands a higher-order framework, such as the self-consistent solution of Dyson equations for both sectors, or the Bethe–Salpeter equation (BSE)~\cite{Rohlfing2000PRB, Onida2002RMP, Strinati1988RNC}, which resums ladder and bubble diagrams to describe excitonic and collective effects. BSE-type methods have been widely applied in solid-state physics to compute optical spectra and polarization response functions, and are well-suited for extending to light–matter coupled systems. Alternatively, nonequilibrium field-theoretic techniques—such as real-time Keldysh approaches~\cite{Stefanucci2013, kamenev2011} or functional renormalization group methods—could be employed to describe the dynamical interplay between cavity photons and correlated electrons. Developing such a framework remains an important direction for future work, particularly for exploring regimes beyond perturbation theory, or for analyzing collective excitations and many-body instabilities induced by cavity coupling.

\section{Photon--Photon Interaction Vertex \( \Gamma^{(4)} \)}

The emergence of photon--photon entanglement in our model can be understood by considering the transformation of an initial two-photon state under the influence of the interaction vertex. Suppose the input state is a separable product of two spectral amplitudes: \( \phi(\omega_1)\phi(\omega_2) \), such as a Fock state of two photons with well-defined frequencies. The output wavefunction, in the absence of loss and under a linear optical process, would remain separable. However, in the presence of an effective interaction mediated by the material system, the outgoing state can acquire non-separable frequency correlations, becoming spectrally entangled.

Formally, the two-photon output wavefunction is given by:
\begin{equation}
\psi_{\mathrm{out}}(\omega_3, \omega_4) = \iint d\omega_1 d\omega_2\; G_4(\omega_3, \omega_4; \omega_1, \omega_2)\, \phi(\omega_1)\, \phi(\omega_2),
\end{equation}
where \( G_4 \) is the full four-point Green’s function, incorporating both disconnected and interacting contributions
and $\phi(\omega)$ is the 
input photon state. 
The lowest-order non-separable term in this expansion is the irreducible vertex \( \Gamma^{(4)}(\omega_3, \omega_4; \omega_1, \omega_2) \), which captures the correlated scattering amplitude.

In our system, energy conservation from the 
self-energy term constrains the process to be elastic, such that \( \omega_1 = \omega_3 \) and \( \omega_2 = \omega_4 \). We therefore write the vertex as a function of the input photon frequencies:
\begin{equation}
\Gamma^{(4)}(\omega_1, \omega_2) \equiv \Gamma^{(4)}(\omega_1, \omega_2; \omega_1, \omega_2).
\end{equation}
Even under this elastic constraint, the structure of \( \Gamma^{(4)}(\omega_1, \omega_2) \) may induce entanglement in the output state. That is,
\begin{equation}
\psi_{\mathrm{out}}(\omega_1, \omega_2) = \Gamma^{(4)}(\omega_1, \omega_2) \phi(\omega_1) \phi(\omega_2)
\end{equation}
is generally non-separable, even if the input \( \phi(\omega) \) is separable.

To quantify this, we now evaluate the structure of \( \Gamma^{(4)}(\omega_1, \omega_2) \) in our model. The interaction vertex arises at fourth order in the polarization expansion and represents the lowest-order diagram in which two photons interact via a pair of virtual polarization loops connected by an electron--hole interaction. 

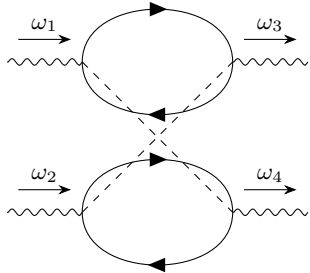
\begin{figure}[h]
\centering
\begin{tikzpicture}
\begin{feynman}
\vertex (i1) at (-2, 1);
\vertex (i2) at (-2, -1);
\vertex (o1) at (2, 1);
\vertex (o2) at (2, -1);
\vertex (vL1) at (-1, 1);
\vertex (vR1) at (1, 1);
\vertex (vL2) at (-1, -1);
\vertex (vR2) at (1, -1);
\diagram* {
  (i1) -- [photon, momentum=\(\omega_1\)] (vL1),
  (i2) -- [photon, momentum=\(\omega_2\)] (vL2),
  (vL1) -- [fermion, half left, looseness=1.2] (vR1),
  (vL2) -- [fermion, half left, looseness=1.2] (vR2),
  (vR1) -- [fermion, half left, looseness=1.2] (vL1),
  (vR2) -- [fermion, half left, looseness=1.2] (vL2),
  (vR1) -- [photon, momentum=\(\omega_3\)] (o1),
  (vR2) -- [photon, momentum=\(\omega_4\)] (o2),
  (vL1) -- [dashed] (vR2),
  (vL2) -- [dashed] (vR1),
};
\end{feynman}
\end{tikzpicture}
\caption{Fourth-order photon–photon scattering diagram via two polarization bubbles coupled by an electron–hole interaction, $V(q,q')$ represented by the dashed lines connecting the two bubbles. 
}
\label{fig:Gamma4}
\end{figure}

To account for interactions between photons mediated by the electronic medium, we consider the fourth-order vertex function \( \Gamma^{(4)}(\omega_1, \omega_2) \), which corresponds diagrammatically to two polarization bubbles coupled via an effective electron–hole interaction potential \( V(k, k') \). The diagram is shown in Fig.~\ref{fig:Gamma4}. This process is the analog of the box diagram in QED that governs light-by-light scattering, but here arises due to interband transitions in a topological insulator.

\begin{widetext}
The corresponding analytical expression is:

\begin{equation}
\Gamma^{(4)}(\omega_1, \omega_2) = \int_{\text{BZ}} \frac{dk}{2\pi} \int_{\text{BZ}} \frac{dk'}{2\pi}
\frac{|\mu(k)|^2}{\omega_1 - \Delta(k) + i\eta}
V(k,k')
\frac{|\mu(k')|^2}{\omega_2 - \Delta(k') + i\eta}.
\label{eq:Gamma4}
\end{equation}

Here, \( \mu(k) \) is the interband transition dipole [Eq.~\eqref{eq:dipole}], and \( V(k,k') \) represents the effective interaction between electron–hole pairs at momenta \( k \) and \( k' \). 
To model the effective photon--photon interaction mediated by virtual excitations in the SSH chain, we adopt a Gaussian kernel of the form
\begin{equation}
V(k, k') = V_0 \exp[-\zeta(k - k')^2],
\end{equation}
which corresponds, in real space, to a finite-range interaction with characteristic length scale \( \ell \sim \zeta^{1/2} \). This form has been widely used in semiconductor quantum optics to model nonlocal exchange interactions and coherence effects in exciton--exciton and polariton--polariton systems~\cite{Ciuti2005,Verger2006,Kira2006,Zimmermann2008}. The parameter \( \zeta \) controls the degree of momentum-space correlation between virtual electron--hole pairs and plays a central role in shaping the biphoton vertex. Specifically, it sets the width of the interaction in momentum space, with larger \( \zeta \) favoring interactions between electronic transitions with nearly equal momenta. 

In the Appendix, we evaluate the 
integrals under the stationary-phase approximation, 
resulting in 
\begin{equation}
\Gamma^{(4)}(\omega_1, \omega_2) \propto 
\frac{ V_0}{(\omega_1 - \Delta_0 + i\eta)(\omega_2 - \Delta_0 + i\eta)} 
\cdot (q^* q'^*)^2 
\cdot \exp\left[ -\zeta (q^* - q'^*)^2 \right] 
\cdot \sqrt{ \frac{2\pi}{\zeta} }.
\end{equation}
\end{widetext}
where 
 \( (q^*, q'^*) \) are the saddle points the integrand. In the limit where the Gaussian interaction kernel is sharply peaked (\( \zeta \to \infty \)), the coupling terms dominate and enforce \(q\approx q' \). 
 Consequently, the saddle points are given by:
 \begin{equation}
 q^* = \sqrt{\frac{2(\omega_1 - \Delta_0)}{\Delta''}}, \qquad
 q'^* = \sqrt{\frac{2(\omega_2 -\Delta_0)}{\Delta''}}.
 \end{equation}

 These expressions identify the momentum components of the electronic transitions that resonate with the incoming photon frequencies. Inserting these saddle points into the exponent \( \Phi(q, q') \) gives the dominant contribution to the integral, and the full vertex function can be approximated accordingly by Gaussian integration around this point.  Both $q^*$ and $q'^*$ must be real
valued.  This sets a threshold condition for the creation 
of the polarization bubbles that 
cause the photon/photon interactions.  This is similar
to the relativistic condition for $e^+/e^-$ pair-production in the vacuum in that the photon energy must be at least $2m_ec^2 
\approx 1.2$MeV. Here, the threshold is set by the SSH energy gap, which can be on the order of 10-100 meV depending on the material. 

Physically,  \( \zeta \), which governs the range of momentum-space correlations, acts as a squeezing parameter for the frequency entanglement of the output state. Larger \( \zeta \) corresponds to stronger suppression of frequency differences (tighter anti-correlations), leading to increased spectral entanglement. In this way, the biphoton vertex transforms a separable input state into a frequency-squeezed, entangled output.

\begin{figure*}[t]
\centering
\begin{tabular}{ccc}
\includegraphics[width=0.28\textwidth]{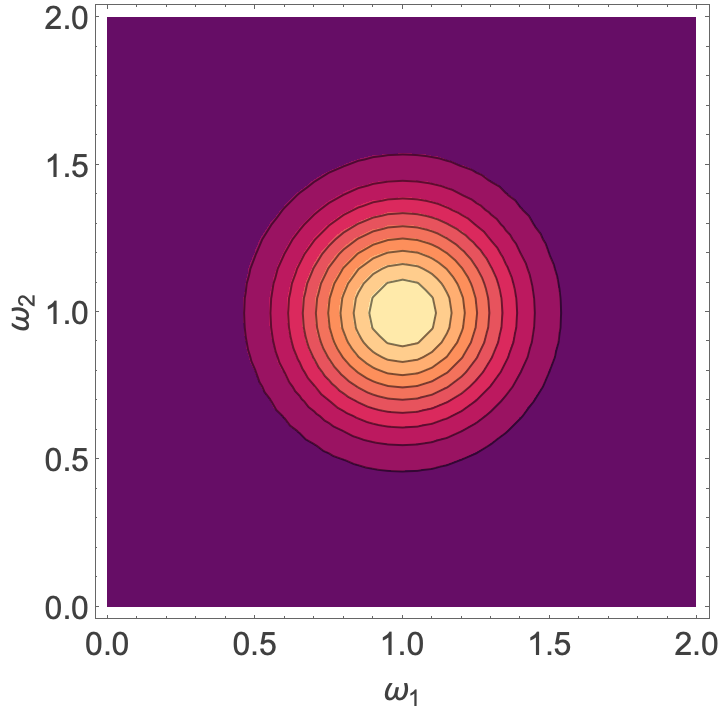} &
\includegraphics[width=0.28\textwidth]{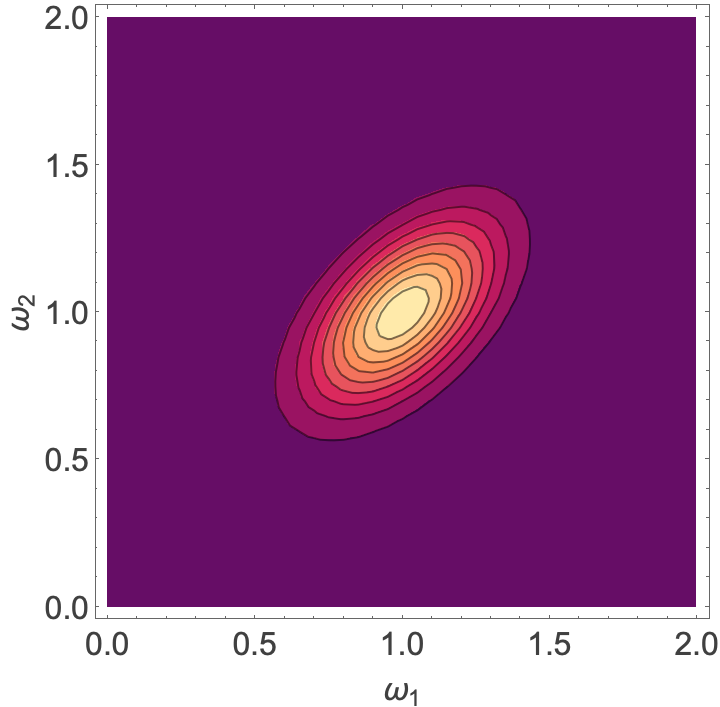} &
\includegraphics[width=0.28\textwidth]{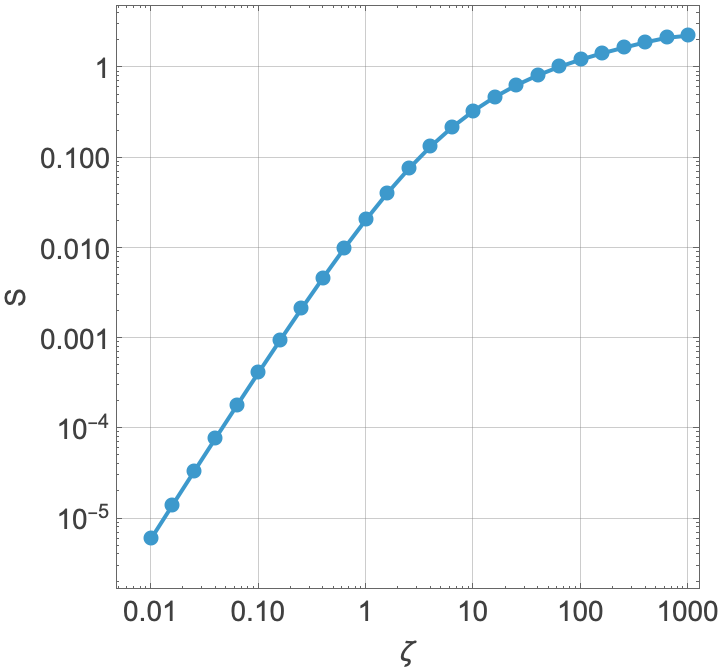} \\
\textbf{(A)} Separable input state & 
\textbf{(B)} Entangled output state ($\zeta\ne 0$) & 
\textbf{(C)} Entanglement entropy $S$ vs. $\zeta$
\end{tabular}
\caption{
Illustration of spectral entanglement induced by the fourth-order interaction vertex $\Gamma^{(4)}(\omega_1, \omega_2)$ acting on a separable biphoton input state. 
\textbf{(A)} The input state $\psi_{\text{in}}(\omega_1, \omega_2) = \phi(\omega_1)\phi(\omega_2)$ is a product of Gaussians centered at $\omega_0 = 1$.
\textbf{(B)} The output state $\psi_{\text{out}}(\omega_1, \omega_2) = \Gamma^{(4)}(\omega_1, \omega_2)\psi_{\text{in}}(\omega_1, \omega_2)$ exhibits frequency anti-correlation due to the nonlocal structure of $\Gamma^{(4)}$.
\textbf{(C)} Entanglement entropy $S$ as a function of interaction range parameter $\zeta$, showing the transition from separable ($S \approx 0$) to highly entangled output. This establishes $\zeta$ as an effective squeezing parameter.
}
\label{fig:entanglement_squeezing}
\end{figure*}

To illustrate how the fourth-order interaction vertex $\Gamma^{(4)}(\omega_1, \omega_2)$ induces spectral entanglement, we consider the transformation of a separable biphoton input state into a correlated output. As shown in Fig.~\ref{fig:entanglement_squeezing}(A), the input state is a product of two Gaussians centered at $\omega_0 = 1$, representing an unentangled, factorable spectral profile. When acted upon by the Gaussian-shaped interaction kernel $\Gamma^{(4)}$, the resulting output state $\psi_{\text{out}}(\omega_1, \omega_2) = \Gamma^{(4)}(\omega_1, \omega_2)\psi_{\text{in}}(\omega_1, \omega_2)$ exhibits strong spectral anti-correlations, as seen in Fig.~\ref{fig:entanglement_squeezing}(B), indicating the emergence of non-separability in the frequency domain.

To quantify the degree of entanglement, we compute the von Neumann entropy of the reduced single-photon density matrix, obtained via Schmidt decomposition of $\psi_{\text{out}}(\omega_1, \omega_2)$. As plotted in Fig.~\ref{fig:entanglement_squeezing}(C), the entanglement entropy $S$ increases monotonically with the interaction range parameter $\zeta$, approaching unity for large $\zeta$. This behavior confirms that $\zeta$ serves as an effective spectral squeezing parameter, modulating the bandwidth over which frequency correlations are imposed by the interaction vertex. In the limit $\zeta \to 0$, the vertex becomes flat and the output state remains separable; in the limit $\zeta \to \infty$, the output state becomes highly entangled, constrained near the diagonal $\omega_1 = \omega_2$.
\subsection{Schmidt Decomposition of the Gaussian Kernel}

To analyze the entanglement properties of the four-photon vertex $\Gamma^{(4)}(\omega_1, \omega_2)$, we consider the simplified form of the kernel arising from the polarization diagram:
\begin{equation}
\Gamma^{(4)}(\omega_1, \omega_2) \propto \exp\left[ -\zeta (q^*(\omega_1) - q^*(\omega_2))^2 \right],
\end{equation}
where $q^*(\omega)$ is the stationary phase point corresponding to photon frequency $\omega$ as identified above.

This symmetric, positive-definite kernel admits a Schmidt decomposition:
\begin{equation}
\Gamma^{(4)}(\omega_1, \omega_2) = \sum_{n=0}^\infty \lambda_n \, u_n(\omega_1) \, u_n(\omega_2),
\end{equation}
with orthonormal Schmidt modes $u_n(\omega)$ related to Hermite functions:
\begin{equation}
u_n(\omega) = \psi_n\big( q^*(\omega) \big), \quad \psi_n(q) = \frac{1}{\sqrt{2^n n! \sqrt{\pi}}} \, e^{-q^2/2} H_n(q).
\end{equation}

The Schmidt coefficients are given by:
\begin{equation}
\lambda_n = \lambda_0 \left( \frac{\zeta}{1 + \zeta} \right)^n,
\end{equation}
with the leading coefficient
\begin{equation}
\lambda_0 = \sqrt{ \frac{2\zeta (1+\zeta)^2}{\pi \left((1+\zeta)^2 + \zeta^2 \right)} }
\end{equation}

Interestingly, the spectrum of Schmidt coefficients $\lambda_n$ depends only on the interaction range parameter $\zeta$, and not on the curvature or topological character of the electronic band structure. This implies that, within the Gaussian approximation, the degree of entanglement between scattered photons is governed primarily by the spatial coherence of the mediating virtual excitations.
The spectral entanglement of the biphoton output can be experimentally accessed through Hong–Ou–Mandel–type interference or frequency-resolved correlation measurements as 
we suggested in our earlier work\cite{Li2019}.

Although the photon self-energy and Kerr nonlinearity show a clear dependence on the topological phase of the SSH chain, our analysis indicates that the four-point photon--photon correlator \( \Gamma^{(4)}(\omega_1, \omega_2) \) is far less sensitive to this topology. The stationary phase evaluation reveals that \( \Gamma^{(4)} \) is dominated by the local curvature of the band structure and the momentum-space correlation kernel \( \zeta \), rather than any global topological invariant such as the Zak phase. Since the vertex depends on correlated virtual transitions at momenta \( k \) and \( k' \), it reflects the spectral geometry of the band edges rather than their winding properties. This outcome is somewhat unexpected: because \( \Gamma^{(4)} \) involves momentum-resolved electron--hole coherence, we had anticipated a more direct signature of topology in the biphoton scattering amplitude. Instead, we find that spectral entanglement and nonlinearity are primarily governed by local band geometry, with only a weak or indirect imprint of the topological phase. This highlights a subtle but important distinction between topological control over linear response functions and its diminished role in shaping multi-photon correlations within a translationally invariant bulk system.
\section{Discussion and Outlook}

In this work, we have developed a microscopic, diagrammatic theory for photon--photon interactions mediated by a topological one-dimensional insulator embedded in an optical microcavity. Using the SSH model as our testbed, we derived the effective fourth-order interaction vertex \( \Gamma^{(4)}(\omega_1, \omega_2) \), which governs biphoton scattering processes within the cavity. The analysis connects topology to optical nonlinearity primarily through the band-dependent dipole matrix elements and the band curvature. However, the four-point vertex \( \Gamma^{(4)} \), which governs biphoton scattering, depends more strongly on local geometric features than on the global topological phase.

While our formalism focuses on bulk-mediated photon--photon interactions within a translationally invariant SSH chain, we note that finite-size effects and edge states can become relevant in experimentally realizable systems. The SSH model under open boundary conditions supports zero-energy edge states in the topological phase; however, these states are localized in real space and do not contribute significantly to the bulk optical response computed in momentum space. In particular, the velocity-gauge light--matter coupling employed here is mediated via interband current matrix elements between extended Bloch states, which do not strongly couple to spatially localized edge modes.

Moreover, in a planar microcavity geometry with uniform mode profiles, the overlap between cavity photons and localized edge states is typically negligible. As a result, the dominant contribution to the nonlinear vertex arises from bulk transitions near the band edge, where the curvature and dipole structure of the Bloch bands govern the Kerr nonlinearity and spectral entanglement. Nonetheless, exploring the role of edge-localized photon--matter interactions in real-space formulations remains an intriguing direction for future work, particularly in nanostructured or few-site implementations where spatial inhomogeneity cannot be neglected.

This bulk-driven nonlinear response is consistent with our prior findings~\cite{Bittner2025jpcl} in a cavity-free setting, where we demonstrated that the third-order DC current response in the SSH model exhibits a \(\pi/2\) phase shift as the system is tuned across the topological transition. That result---obtained in the absence of any cavity or boundary-localized states---confirms that the nonlinear phase behavior is a fingerprint of the bulk band geometry. In the present context, the photon self-energy inherits these topological features through the momentum-resolved dipole matrix elements and curvature of the SSH bands, while the fourth-order scattering vertex reflects primarily the curvature and local coherence of the bands, with only weak sensitivity to topology. Exploring the interplay between bulk and edge effects in confined or nanostructured geometries remains an important direction for future work.

Several key results emerge from this framework:
\begin{itemize}
  \item The photon self-energy \( \Sigma(\omega) \) inherits topological features of the SSH chain through the structure of \( \mu(k)^2 \) and the gap function \( \Delta(k) \). The threshold for absorption, its slope, and the spectral density all vary significantly with the hopping ratio \( r = t_2/t_1 \).

  \item The effective Kerr nonlinearity \( U \sim \Gamma^{(4)}(\omega, \omega) \) can be tuned in both magnitude and sign across the topological phase transition. This gives rise to a ``nonlinear topological phase diagram'' in which \( U > 0 \) or \( U < 0 \), with singularities emerging at band-gap closures.

  \item The biphoton vertex exhibits spectral entanglement controlled by the curvature \( \Delta'' \) and interaction range \( \zeta \). This structure is shaped by local geometric properties of the bands rather than global topological invariants. In the flat-band limit, the vertex becomes separable and yields unentangled photons; conversely, in the presence of curvature, frequency entanglement emerges as a direct fingerprint of the underlying band geometry.

  \item Our stationary phase analysis links entanglement and nonlinearity to analytic properties of the integrand, establishing a predictive link between curvature, dipole distribution, and quantum optical response.
\end{itemize}

The diagrammatic techniques developed here provide a natural foundation for generalization. By replacing the SSH model with other topological media---such as Haldane chains, Su--Schrieffer--Heeger ladders, or higher-dimensional Chern insulators---one can investigate topologically enhanced photon blockade, optical bistability, and spontaneous parametric downconversion in strongly correlated photonic systems. The analogy with vacuum QED is particularly striking: whereas photon--photon scattering in standard quantum electrodynamics arises from virtual electron--positron pairs, its condensed matter counterpart emerges from virtual electron--hole excitations within a solid-state medium, yielding an analogous polarization bubble structure now endowed with topological control.

{\color{black}
The proposed theoretical framework of photon–photon scattering can be further developed to enable both control and probing of different phases of matter, as well as phase transitions in the emerging field of quantum materials \cite{Schlawin:2022aa}. In particular, for the case of cavity superconductivity, one can consider the variation of entanglement entropy arising from biphoton scattering by Cooper pairs bound through cavity photons or exciton–polaritons. Moreover, extending the scattering approach to higher-dimensional Chern insulators, and incorporating chiral light, is essential, as it may significantly enhance the sensitivity of entanglement entropy to interactions with topological states.

Conversely, in low-dimensional materials, light can be confined to form waveguide or cavity modes that self-hybridize with electronic excitations (excitons) or phonons, resulting in exciton/phonon–polariton states \cite{CanalesJCP:2021}. Notably, self-hybridized exciton–polaritons have been recently observed in the van der Waals magnetic semiconductor CrSBr \cite{RutaNatComm:2023}, exhibiting signatures of the ultra-strong coupling regime \cite{DirnbergerNature:2023}. Similarly, self-hybridized phonon–polaritons have been investigated in hBN \cite{kumarNanoLet:2015}. Applying our diagrammatic technique to photon–photon scattering in these systems can provide valuable insights into bipolariton dynamics in this emerging class of cavity quantum materials.
}

Our calculations yield an effective Kerr interaction with negative sign ($U < 0$), indicating an attractive photon–photon interaction mediated by virtual excitations of the SSH chain. This result implies a focusing-type nonlinearity that supports bright soliton formation and polariton condensation in the cavity field. Similar behavior has been observed experimentally in polaritonic systems where bright solitons emerge under conditions of strong coupling and low loss~\cite{Liew2010PRL, Amo2011Science}. These phenomena require an underlying attractive nonlinearity, consistent with our prediction. However, direct experimental determination of the sign of $U$ remains subtle; most evidence is indirect and inferred from the observation of self-localized modes, condensate profiles, or nonlinear shifts~\cite{Ballarini2017NanoLett, Estrecho2021PRL}. Within our model, the sign and magnitude of $U$ arise from the curvature and dipole structure of the Bloch bands and are strongly momentum-selective. The fact that $U < 0$ emerges naturally across a broad parameter range reinforces the idea that topological materials embedded in cavity QED architectures can serve as tunable nonlinear media with built-in mechanisms for photon attraction and condensation.

In parallel with the photon self-energy and fourth-order vertex analysis, we have also explored how the electronic states themselves are renormalized by vacuum fluctuations of the cavity mode. Using a one-loop Feshbach projection, we derived the effective fermionic self-energy induced by virtual photon exchange and demonstrated that the cavity mediates interband mixing between conduction and valence states. This leads to a renormalized, momentum-resolved electronic dispersion in which the bands hybridize and shift in response to virtual cavity dressing. Unlike molecular systems where cavity coupling is typically weak on a per-molecule basis, the renormalization here extends across the entire Brillouin zone and is inherently collective. These results not only establish a microscopic foundation for vacuum-induced band reshaping, but also bear on broader debates in the literature concerning cavity-modified chemistry and materials design.  However, 
we emphasize that for physically 
reasonable values of the
coupling $g$, this renormalization is 
very small across the entire Brillouin
zone and 
vanishes at the Brillouin zone edge
where the energy gap between 
conduction and valance bands is the smallest and where one would 
expect the consequences of cavity-induced band
mixing to be most important.

These findings resonate with ongoing debates in polaritonic chemistry, where cavity-induced modifications of ground-state reactivity have been reported in molecular ensembles~\cite{Hirai2023ChemRev, Mandal2023ChemRev}. While some studies attribute such effects to collective strong coupling or vibrational polariton formation~\cite{Flick2024CommMat, CamposGonzalezAngulo2024PRR}, others caution that symmetry, disorder, and mode selectivity can suppress reactivity shifts~\cite{Herrera2025ChemSocRev}. Our results reinforce this view: despite coherent coupling across the Brillouin zone, the vacuum field fails to significantly reorganize the electronic states near the Fermi level when dipole coupling vanishes by symmetry. This suggests that care must be taken in attributing chemical effects to equilibrium cavity dressing alone.

A more complete understanding will require a fully self-consistent framework that captures the mutual dressing of photons and electrons. Techniques such as Bethe–Salpeter ladder resummation~\cite{Rohlfing2000PRB, Onida2002RMP} or nonequilibrium Keldysh field theory~\cite{Stefanucci2013,kamenev2011} provide natural directions for extending the present formalism beyond the perturbative regime, and toward the analysis of cavity-induced instabilities, polariton condensation, or topological phase engineering in driven quantum materials.

This correspondence opens several promising directions for future development. One avenue involves non-perturbative extensions, such as summing ladder and bubble diagrams to all orders or resumming the full Bethe--Salpeter kernel to study cavity-bound states of interacting photons. Another is to draw analogies with quantum impurity physics by treating the cavity photon as an impurity coupled to a correlated topological bath, enabling the use of Keldysh or Matsubara techniques to compute spectral functions. The formalism may also be extended to time-resolved and pump--probe spectroscopies, providing access to transient nonlinear response and four-wave mixing signals~\cite{Bittner2025jpcl}. Finally, incorporating multimode cavity fields---spatially structured or in extended arrays---could enable simulations of polariton condensation and entanglement in driven, topologically nontrivial photonic systems.

An intriguing consequence of our results is the emergence of an effective Kerr interaction with negative sign, $U < 0$, across both the trivial and topological phases. This sign indicates an attractive photon--photon interaction, such that the cavity field satisfies a focusing-type nonlinearity. In the semiclassical limit, the dynamics of the cavity mode can be described by a Gross--Pitaevskii-type equation for the condensate wavefunction $\psi(t) \sim \langle a(t) \rangle$, given by
\begin{equation}
i \frac{\partial \psi}{\partial t} = \omega_0 \psi + U |\psi|^2 \psi.
\end{equation}
For $U < 0$, this equation supports bright soliton solutions, in which the photon field becomes spatially localized and self-trapped due to the attractive nonlinearity. Notably, in our model, the origin of this effective attraction lies not in direct photon--photon interactions, but rather in the nonperturbative dressing of the cavity mode by virtual electronic excitations of the SSH medium. The topology of the underlying band structure shapes the spectral profile and spatial coherence of this dressing, which in turn influences the sign and magnitude of $U$. This result opens the possibility of engineering topologically controlled photonic solitons and nonlinear quantum states of light through materials-based design of the electronic environment within an optical cavity.

While the SSH model undergoes a well-defined topological phase transition at the critical point $t_1 = t_2$, the specific values $t_2 = 0.5\, t_1$ and $t_2 = 1.5\, t_1$ serve as important benchmark regimes for examining the physics of the trivial and topological phases, respectively. In both cases, the energy gap remains sizable, $\Delta_{\mathrm{gap}} = 2|t_1 - t_2| = 1$, ensuring spectral isolation of the conduction and valence bands. At $t_2 = 0.5\, t_1$, the system lies deep in the trivial phase with strongly localized Wannier orbitals centered on the $t_1$-bonds and a Zak phase of zero. Conversely, for $t_2 = 1.5\, t_1$, the system is well within the topological regime, characterized by a Zak phase of $\pi$, inversion of Wannier center positions, and the emergence of topologically protected edge states under open boundary conditions. These points thus provide clean, non-critical settings in which the nonlinear optical response and cavity-induced interactions can be probed, while retaining the topological distinctions encoded in the underlying band structure and Berry curvature.

These findings parallel our recent results on the nonlinear optical response of topological systems computed within a velocity-gauge formalism. In our prior work on the SSH model, we demonstrated that the third-order current response function \( \chi^{(3)}(\omega_3, \tau_2, \omega_1) \) exhibits a distinct topological phase signature: a \(\pi/2\) rotation of the complex-valued DC response as the system is tuned across the topological transition point \( t_1 = t_2 \)~\cite{Bittner2025jpcl}. This nonlinear fingerprint arises from phase-sensitive quantum coherence pathways and reflects changes in the geometric structure of the Bloch wavefunctions---particularly the winding of the Berry connection. The emergence of such phase inversions in the absence of edge states or symmetry breaking underscores the capacity of nonlinear spectroscopy to diagnose topological order in bulk materials.

Ultimately, this work points toward a new class of nonlinear quantum optical materials in which topology governs linear transport and nonlinear optical response, while multi-photon coherence and entanglement depend more subtly on band geometry than on topological phase.

\begin{acknowledgments}
The work at the University of Houston was supported by the National Science Foundation under CHE-2404788 and the Robert A. Welch Foundation (E-1337). A.P. research was supported by the Laboratory Directed Research and Development program of Los Alamos National Laboratory under project number 20230347ER. This work was performed, in part, at the Center for Integrated Nanotechnologies, an Office of Science User Facility operated for the U.S. Department of Energy (DOE) Office of Science. Los Alamos National Laboratory, an affirmative action equal opportunity employer, is managed by Triad National Security, LLC for the U.S. Department of Energy’s NNSA, under contract 89233218CNA000001.


\end{acknowledgments}

\section*{Data Accessibility Statement.}
The authors declare that the data supporting the findings of this study are available within the paper.

\appendix

\section{An Overview of the SSH Model}

The Su-Schrieffer-Heeger (SSH) model is a paradigmatic one-dimensional tight-binding model exhibiting a topological phase transition driven by dimerization \cite{SSH1979}. It consists of a bipartite chain with alternating hopping amplitudes \( t_1 \) and \( t_2 \), connecting adjacent lattice sites of sublattices \( A \) and \( B \). The model Hamiltonian in real space is:
\begin{equation}
H_{\text{SSH}} = \sum_n \left( t_1 c^\dagger_{n,A} c_{n,B} + t_2 c^\dagger_{n+1,A} c_{n,B} + \text{h.c.} \right),
\end{equation}
where \( c^\dagger_{n,\alpha} \) creates an electron at unit cell \( n \) on sublattice \( \alpha \in \{A,B\} \).
Transforming to momentum space using the basis \( \Psi_k = (c_{k,A}, c_{k,B})^T \), the Hamiltonian becomes:
\begin{equation}
H_{\text{SSH}}(k) = \begin{pmatrix}
0 & h(k) \\
h^*(k) & 0
\end{pmatrix}, \quad
h(k) = t_1 + t_2 e^{-ik}.
\end{equation}
The energy bands are given by:
\begin{equation}
\varepsilon_{\pm}(k) = \pm\frac{1}{2} \Delta(k), \quad
\Delta(k) =2 \sqrt{t_1^2 + t_2^2 + 2t_1 t_2 \cos k}.
\label{eq:SSH_gap}
\end{equation}
Here, $\Delta(k)$ corresponds to the energy to create a single electron/hole excition at $k$. 
The bandgap  closes at \( k = \pi \) when \( t_1 = t_2 \), marking the topological transition point. For \( t_2 > t_1 \), the system is in the nontrivial topological phase, characterized by a Zak phase of \( \pi \) and the presence of zero-energy edge states under open boundary conditions.
The corresponding eigenstates of the lower (valence) and upper (conduction) bands are:
\begin{align}
\ket{u_v(k)} &= \frac{1}{\sqrt{2}} \begin{pmatrix}
- e^{-i \theta(k)} \\
1
\end{pmatrix}, \quad
\ket{u_c(k)} = \frac{1}{\sqrt{2}} \begin{pmatrix}
e^{-i \theta(k)} \\
1
\end{pmatrix},
\end{align}
where \( \theta(k) = \arg(h(k)) \).


\begin{widetext}
\section{Stationary Phase Evaluation of $\Gamma^{(4)}$}

An instructive and technically elegant approach to evaluating the fourth-order scattering vertex is to recast the full integrand as a single exponential, combining all multiplicative structure into a logarithm. This allows one to systematically apply the two-dimensional stationary phase method using a single scalar-valued exponent.

We begin with the expression for the interaction vertex:
\begin{equation}
\Gamma^{(4)}(\omega_1, \omega_2) = V_0 \int dq \, dq'\; 
\frac{q^2 q'^2 \, e^{-\zeta(q - q')^2}}{[\omega_1 - \Delta(q) + i\eta][\omega_2 - \Delta(q') + i\eta]}.
\end{equation}

Near the band edge at \( k = \pi \), we expand the gap as:
\begin{equation}
\Delta(q) = \Delta_0 + \tfrac{1}{2} \Delta'' q^2,
\end{equation}
and approximate the transition dipole amplitude as \( \mu(q) \approx A q \). We define the integrand as:
\begin{equation}
I(q, q') = \frac{q^2 q'^2 \, e^{-\zeta(q - q')^2}}{[\omega_1 - \Delta(q) + i\eta][\omega_2 - \Delta(q') + i\eta]}.
\end{equation}

Then write the entire integrand as an exponential:
\begin{equation}
\Gamma^{(4)}(\omega_1, \omega_2) = V_0 \int dq \, dq' \; e^{\Phi(q, q')},
\end{equation}
where the exponent is
\begin{align}
\Phi(q, q') 
&= \log(q^2 q'^2) - \zeta(q - q')^2 
- \log[\omega_1 - \Delta(q) + i\eta] - \log[\omega_2 - \Delta(q') + i\eta] \notag \\
&= \log(q^2 q'^2) - \zeta(q - q')^2 
- \log\left[\omega_1 - \Delta_0 - \tfrac{1}{2} \Delta'' q^2 + i\eta\right] 
- \log\left[\omega_2 - \Delta_0 - \tfrac{1}{2} \Delta'' q'^2 + i\eta\right].
\end{align}

The saddle points \( (q^*, q'^*) \) are obtained by solving the stationary phase conditions:
\begin{equation}
\frac{\partial \Phi}{\partial q} = 0, \qquad \frac{\partial \Phi}{\partial q'} = 0.
\end{equation}

Computing these derivatives:
\begin{align}
\frac{\partial \Phi}{\partial q} 
&= \frac{2}{q} - 2\zeta(q - q') 
+ \frac{\Delta'' q}{\omega_1 - \Delta_0 - \tfrac{1}{2} \Delta'' q^2 + i\eta} = 0, \\
\frac{\partial \Phi}{\partial q'} 
&= \frac{2}{q'} + 2\zeta(q - q') 
+ \frac{\Delta'' q'}{\omega_2 - \Delta_0 - \tfrac{1}{2} \Delta'' q'^2 + i\eta} = 0.
\end{align}

In the limit where the Gaussian kernel is narrow (\( \zeta \) large), the dominant contribution comes from near \( q \approx q' \), and the second terms cancel in the two equations. Neglecting imaginary corrections and solving for resonance conditions, we find:
\begin{equation}
q^* = \sqrt{\frac{2(\omega_1 - \Delta_0)}{\Delta''}}, \qquad
q'^* = \sqrt{\frac{2(\omega_2 - \Delta_0)}{\Delta''}}.
\end{equation}

We now expand \( \Phi(q, q') \) to second order around \( (q^*, q'^*) \) and approximate the integral as a Gaussian over \( (q, q') \):
\begin{equation}
\Gamma^{(4)}(\omega_1, \omega_2) \approx V_0 \, e^{\Phi(q^*, q'^*)} \cdot \sqrt{\frac{(2\pi)^2}{\det H}},
\end{equation}
where \( H \) is the Hessian matrix of second derivatives \( \partial^2 \Phi/\partial q_i \partial q_j \) evaluated at the stationary point. The leading contribution comes from:
\begin{equation}
\Phi(q^*, q'^*) = \log(q^{*2} q'^{*2}) - \zeta(q^* - q'^*)^2 
- \log[\omega_1 - \Delta_0 - \tfrac{1}{2} \Delta'' q^{*2} + i\eta] 
- \log[\omega_2 - \Delta_0 - \tfrac{1}{2} \Delta'' q'^{*2} + i\eta].
\end{equation}

Using the resonance condition \( \omega_i - \Delta_0 = \tfrac{1}{2} \Delta'' q_i^{*2} \), the final expression simplifies to:
\begin{equation}
\Gamma^{(4)}(\omega_1, \omega_2) \approx 
\frac{A^4 V_0}{(\omega_1 - \Delta_0 + i\eta)(\omega_2 - \Delta_0 + i\eta)} 
\cdot (q^* q'^*)^2 
\cdot \exp\left[ -\zeta (q^* - q'^*)^2 \right] 
\cdot \sqrt{ \frac{2\pi}{\zeta} }.
\end{equation}
This formulation offers a compact path to the saddle-point approximation and highlights the Gaussian entanglement in energy-momentum space introduced by the interaction kernel \( e^{-\zeta(q - q')^2} \).

\section{Feshbach Projection and the Effective Action for the Photon Field}

In this section, we provide a more detailed derivation of the effective photon Green's function using the Feshbach projection formalism applied to the full system composed of a cavity mode coupled to a gapped electronic medium (e.g., the SSH chain). Our goal is to derive an effective quadratic action for the photon field by integrating out the electronic degrees of freedom, and to show that this result is formally equivalent to the Dyson equation obtained from the Keldysh path-integral formalism.

We begin with the total system Hamiltonian:
\begin{equation}
H = H_{\mathrm{SSH}} + \omega_c a^\dagger a + g \sum_k \mu(k)\left[ a c_k^\dagger v_k + a^\dagger v_k^\dagger c_k \right],
\end{equation}
where $H_{\mathrm{SSH}}$ describes the electronic SSH chain, $a$ is the photon annihilation operator for the cavity mode of frequency $\omega_c$, and the light--matter interaction term is linear in the photon field and couples interband transitions between conduction ($c_k$) and valence ($v_k$) bands via dipole matrix elements $\mu(k)$.

To proceed, we write the full generating functional in terms of coherent-state path integrals:
\begin{equation}
\mathcal{Z} = \int \mathcal{D}[a^*, a] \mathcal{D}[\bar{c}, c] \mathcal{D}[\bar{v}, v] \, e^{i S[a^*, a; \bar{c}, c; \bar{v}, v]},
\end{equation}
where the total action $S$ includes contributions from the photon mode, the electronic bands, and the coupling term.

Assuming the electronic sector is quadratic and initially uncorrelated with the photon field, we can integrate out the fermionic fields exactly. The result is an effective action for the photon mode:
\begin{equation}
S_{\mathrm{eff}}[a^*, a] = S_0[a^*, a] - i \, \mathrm{Tr} \log \left[ 1 + g^2 \, \mathcal{G}_v(\omega) \, \mu(k)^2 \, \mathcal{G}_c(\omega) \, |a(\omega)|^2 \right],
\end{equation}
where $\mathcal{G}_{c,v}(\omega)$ are the bare Green's functions for the conduction and valence bands, and the trace is over both momentum and frequency.

At the one-loop level, expanding the log to quadratic order in $a$ gives:
\begin{equation}
S_{\mathrm{eff}}[a^*, a] = \int \frac{d\omega}{2\pi} \, a^*(\omega) \left[ \omega - \omega_c - \Sigma^R(\omega) + i\eta \right] a(\omega),
\end{equation}
with the retarded photon self-energy:
\begin{equation}
\Sigma^R(\omega) = g^2 \int \frac{dk}{2\pi} \frac{\mu(k)^2}{\omega - \Delta(k) + i\eta},
\end{equation}
where $\Delta(k)$ is the interband excitation energy between valence and conduction bands at momentum $k$. This matches precisely the result obtained from the Dyson equation in Keldysh theory:
\begin{equation}
G^R(\omega) = \left[ \omega - \omega_c - \Sigma^R(\omega) + i\eta \right]^{-1}.
\end{equation}

To describe nonequilibrium dynamics, the full Keldysh Green's function includes both the retarded and Keldysh components:
\begin{equation}
G^K(\omega) = G^R(\omega) \Sigma^K(\omega) G^A(\omega),
\end{equation}
where $\Sigma^K(\omega)$ is the Keldysh component of the photon self-energy, which encodes the statistical properties (occupations) of the bath. In thermal equilibrium, the fluctuation--dissipation theorem relates this to the imaginary part of $\Sigma^R$:
\begin{equation}
\Sigma^K(\omega) = -2i \, \mathrm{Im} \Sigma^R(\omega) \left[1 + 2n_B(\omega) \right],
\end{equation}
with $n_B(\omega) = (e^{\omega/T} - 1)^{-1}$ the Bose distribution. It follows that
\begin{equation}
G^K(\omega) = -2i \, A(\omega) \left[1 + 2n_B(\omega) \right], \quad A(\omega) = -\frac{1}{\pi} \mathrm{Im} G^R(\omega),
\end{equation}
where $A(\omega)$ is the photon spectral function. Thus, the occupation and coherence properties of the photon mode are governed entirely by the imaginary part of the retarded self-energy derived from the Feshbach procedure.

This correspondence validates the use of the Feshbach projection as a systematic and physically transparent tool for deriving the nonlinear optical response of cavity-coupled topological systems. It also provides a clear interpretation of how the topology and band geometry of the underlying electronic medium become encoded in the spectral and dynamical properties of the cavity photon. Moreover, it confirms that the nonequilibrium photon statistics—such as squeezing, thermalization, or entanglement—can be accurately captured using the same self-energy input, without the need to explicitly evolve the full many-body wavefunction of the bath.

\section{Explicit Evaluation of the Keldysh Green's Function}

We now provide a step-by-step derivation of the Keldysh Green's function $G^K(\omega)$ for the cavity photon field in thermal equilibrium, using the self-energy obtained via the Feshbach projection. This demonstrates that the nonequilibrium photon statistics are fully encoded in the retarded self-energy $\Sigma^R(\omega)$.

\subsection{Dyson equation for the retarded propagator}

We begin with the retarded Green's function:
\begin{equation}
G^R(\omega) = \frac{1}{\omega - \omega_c - \Sigma^R(\omega) + i\eta},
\end{equation}
where $\Sigma^R(\omega) = \Delta_R(\omega) - i \Gamma(\omega)$ encodes the real and imaginary parts of the photon self-energy. The imaginary part $\Gamma(\omega) = -\mathrm{Im} \Sigma^R(\omega)$ describes the damping and spectral broadening due to coupling to the electronic bath.

The advanced Green's function is given by complex conjugation:
\begin{equation}
G^A(\omega) = \left[G^R(\omega)\right]^* = \frac{1}{\omega - \omega_c - \Delta_R(\omega) - i(\Gamma(\omega) + \eta)}.
\end{equation}

\subsection{Keldysh component from fluctuation--dissipation relation}

In thermal equilibrium, the Keldysh self-energy is related to the imaginary part of the retarded self-energy via the fluctuation--dissipation theorem:
\begin{equation}
\Sigma^K(\omega) = -2i\, \mathrm{Im} \Sigma^R(\omega) \left[ 1 + 2n_B(\omega) \right] = 2i\, \Gamma(\omega) \left[ 1 + 2n_B(\omega) \right],
\end{equation}
where $n_B(\omega) = (e^{\omega/T} - 1)^{-1}$ is the Bose--Einstein distribution.

The full Keldysh Green's function is then:
\begin{equation}
G^K(\omega) = G^R(\omega) \, \Sigma^K(\omega) \, G^A(\omega) = 2i\, \Gamma(\omega) \left[ 1 + 2n_B(\omega) \right] \cdot |G^R(\omega)|^2.
\end{equation}

\subsection{Spectral representation and physical meaning}

We define the photon spectral function as:
\begin{equation}
A(\omega) = -\frac{1}{\pi} \, \mathrm{Im} G^R(\omega) = \frac{1}{\pi} \, \frac{\Gamma(\omega) + \eta}{\left[ \omega - \omega_c - \Delta_R(\omega) \right]^2 + \left[ \Gamma(\omega) + \eta \right]^2}.
\end{equation}

Thus, the Keldysh Green's function becomes:
\begin{equation}
G^K(\omega) = -2i\, A(\omega) \left[ 1 + 2n_B(\omega) \right].
\end{equation}

This result shows that the occupation and quantum statistical fluctuations of the cavity field are governed entirely by the spectral function $A(\omega)$ and the Bose distribution. In particular, $G^K(\omega)$ contains the same structure as $A(\omega)$, modulated by the thermal occupation factor. This provides a clear and quantitative bridge between the Feshbach-derived $\Sigma^R(\omega)$ and the full nonequilibrium photon statistics captured within the Keldysh formalism.

\end{widetext}

\bibliographystyle{apsrev4-2}
\bibliography{References_local}

\end{document}